\documentclass[twocolumn,prb,superscriptaddress,showpacs,citeautoscript,floatfix]{revtex4}

\usepackage{graphicx}
\usepackage{dcolumn}   % column centering
\usepackage{bm}        % bold math
\usepackage{flafter}
\usepackage{color}
\usepackage{hyperref}

\begin{document}
\title{Pressure-induced suppression of charge density wave and emergence of superconductivity in 1$\textit{T}$-VSe$_2$}

\author{S. Sahoo}
\affiliation{HP$\&$SRPD, Bhabha Atomic Research Centre, Trombay, Mumbai 400085, India}
\affiliation{Department of Physical Sciences, Homi Bhabha National Institute,
Anushaktinagar, Mumbai 400094, India}

\author{U. Dutta}
\affiliation{HP$\&$SRPD, Bhabha Atomic Research Centre, Trombay, Mumbai 400085, India}
\affiliation{Department of Physical Sciences, Homi Bhabha National Institute,
Anushaktinagar, Mumbai 400094, India}

\author{L. Harnagea}
\affiliation{Department of Physics, Indian Institute of Science Education and Research (IISER), Pune 411008, India}

\author{A. K. Sood}
\email[E-mail:~]{asood@iisc.ac.in}
\affiliation{Department of Physics, Indian Institute of Science, Bangalore, 560012, India}

\author{S. Karmakar}
\email[E-mail:~]{sdak@barc.gov.in}
\affiliation{HP$\&$SRPD, Bhabha Atomic Research Centre, Trombay, Mumbai 400085, India}
\affiliation{Department of Physical Sciences, Homi Bhabha National Institute,
Anushaktinagar, Mumbai 400094, India}

\date{\today}

\begin{abstract}

We report pressure evolution of charge density wave (CDW) order and emergence of superconductivity (SC) in 1$\textit{T}$-VSe$_2$ single crystal by studying resistance and magnetoresistance behavior under high pressure. With increasing quasi-hydrostatic pressure the CDW order enhances with increase of the ordering temperature up to 240K at 12 GPa. Upon further increase of pressure, the resistance anomaly due to CDW order gets suppressed drastically and superconductivity emerges at $\sim$15 GPa, with the onset critical temperature (T$_c$) $\sim$4K. The pressure dependence of T$_c$ is found negligible, different from the significant increase or a dome-shape seen in iso-structural layered diselenide superconductors. The high pressure magnetoresistance and Hall measurements suggest successive electronic structural changes with Fermi surface modifications at 6 GPa and 12 GPa. From the observed negative magnetoresistance in this pressure range and absence of coexisting CDW and SC phases, we propose that intra-layer spin-fluctuation can play a role in the emergence of superconductivity in the high pressure phase.
\end{abstract}

%\pacs{PACS: 74.70.Ad, 74.62.Fj, 71.45.Lr, 74.25.Dw}

%\pacs{71.45.Lr, %CDW collective excitations
%     74.25.Dw, %Phase diagram superconductivity
%     74.25.Ha, %Magnetic properties of superconductor
%     74.70.Ad, %SC materials binary compounds
%     71.20.Gj, Electronic structure of Semimetals,
%     74.62.Fj, Pressure effects on SC transition temperature variation,
%     }%

\maketitle

%\section{Introduction}

Layered transition metal dichalcogenide (TMDC) compounds provide an ideal platform to explore exotic ground-state electronic orders by tuning the Fermi surface topology and many-body effects through various external stimulations~\cite{Kolobov2016}. Among these, 1$\textit{T}$-structured correlated metals (e.g., 1$\textit{T}$-TaS$_2$, 1$\textit{T}$-TaSe$_2$, 1$\textit{T}$-TiSe$_2$ and 1$\textit{T}$-TiTe$_2$) have been extensively studied for understanding the mechanism of charge density wave (CDW) order at low temperature and its coexistence with superconductivity (SC) in some part of the phase diagram~\cite{Sipos2008,Ang2012,Wang2017,Morosan2006,Kusmartseva2009,Dutta2018,Dutta2019}. An unconventional (exciton or band-type Jahn-Teller) mechanism for the CDW ordering has been established in these compounds by several experimental and theoretical studies, rather than the conventional Fermi surface nesting mechanism. However, the underlying mechanism for superconductivity is not conclusive so far. In some systems, the dome-shaped superconducting window in the vicinity of the CDW suppressed quantum critical point (QCP) supports unconventional SC scenario where CDW amplitude fluctuation is believed to be responsible for the Cooper pair formation~\cite{Barath2008,Joe2014,Kogar2017}. In some other systems phonon mediated (BCS type) SC appears in the phase diagram window (separated from the CDW region) that is believed to originate in phase-separated metallic domains~\cite{Li2007,Liu2016}.

1$\textit{T}$-VSe$_2$ is one of the rare correlated metallic systems where three dimensional (3D) nesting of the Fermi surface gives rise to 3D-CDW ordering (having commensurate in-plane wave vector 0.25\textbf{a*} with an incommensurate out-of-plane component)~\cite{Tsutsumi1982,Terashima2003,Strocov2012,Jolie2019}. The CDW transition temperature (T$_{CDW}$) is $\sim$110K, as seen in resistivity and susceptibility measurements~\cite{Bruggen1976,Barua2017}. Due to weak nesting condition, the CDW distortion (amplitude) is small, making the superstructure bands not observable below T$_{CDW}$ in ARPES measurements ~\cite{Terashima2003,Strocov2012}. Although the high resolution ARPES measurements on single crystal 1$\textit{T}$-VSe$_2$ show the presence of only hole pocket (of V 3d$_{z^2}$ band) at the M point of the Brillouin zone (BZ) as supported by the DFT calculations, the observed strong hybridization of Se 4p$_{x,y}$ and 4p$_z$ at the $\Gamma$ point near the Fermi level ~\cite{Strocov2012} may turn the system into a multiband character by forming a 4p hole pocket. The Fermi surface topology thus appears to be very much susceptible to external perturbations like intercalation, reduced thickness or pressure to reveal exotic physical properties. While electron doping by alkali metal intercalation is found to alter the Fermi level band structure drastically into a 2D character ~\cite{Starnberg1993,Brauer1998}, reduction of layer thickness shows anomalous change in the CDW ordering temperature due to dimensional crossover, reduced interlayer coupling and enhanced quantum confinement~\cite{Yang2014,Pasztor2017}. The study of monolayer VSe$_2$ has been of tremendous current interest as various synthesis procedures, substrate and strain conditions modify the Fermi surface drastically with emergence of different CDW order with distinct ordering temperatures, energy gap, Fermi arc and Mott/Peierls insulating state ~\cite{Zhang2017,Chen2018,Duvjir2018,Umemoto2019}. Monolayer VSe$_2$ has also been predicted to be energetically close to a spin-ordered structure ~\cite{Esters2017}, which has indeed been recently observed ~\cite{Wong2019}.

In spite of all these exotic physical properties, only very few investigations have so far been reported on the bulk 1$\textit{T}$-VSe$_2$~\cite{Bruggen1976,Friend1978,Barua2017}. This is primarily due to the difficulties in the synthesis of pristine compound. It always grows as V-rich compound (1$\textit{T}$-V$_{1+\delta}$Se$_2$) where extra V atoms are intercalated in the interlayer, giving rise to strong Curie-Weiss (CW) paramagnetic behavior at low temperature. These localized moments also act as Kondo scattering centers ~\cite{Barua2017}, which can also give rise to weak magnetic ordering at low temperature, hindering the study of intrinsic properties of 1$\textit{T}$-VSe$_2$. In this Rapid Communication, we report on the transport properties of 1$\textit{T}$-VSe$_2$ single crystal under high pressure. The in-plane resistance at room temperature shows anomalous pressure variation at $\sim$6 GPa, indicating change in its band structure and possibly the Fermi surface (FS) topology. With increasing pressure CDW ordering is dramatically enhanced up to 12 GPa with ordering temperature reaching up to 240K. At further higher pressures the CDW feature is drastically suppressed and superconducting state emerges above 15 GPa. The SC T$_c$ increases marginally with pressure, T$_c$ reaching $\sim$5K at 22 GPa, the highest pressure of our measurements. The low temperature negative magnetoresistance (MR) due to Kondo scattering of interlayer V atoms is suppressed strongly above 12 GPa, agreeing with corresponding vanishing of resistance upturn. The observed positive MR in the SC phase is typical of similar layered TMDCs. The Hall measurements further helps understand the FS modification under pressure leading to suppression of the CDW phase and emergence of SC. Upon decompression, significant resistance drop due to the onset of SC is observed down to 10 GPa, but without any sign of the coexistent CDW phase.

\begin{figure}[tb]
\centerline{\includegraphics[width=90mm,clip]{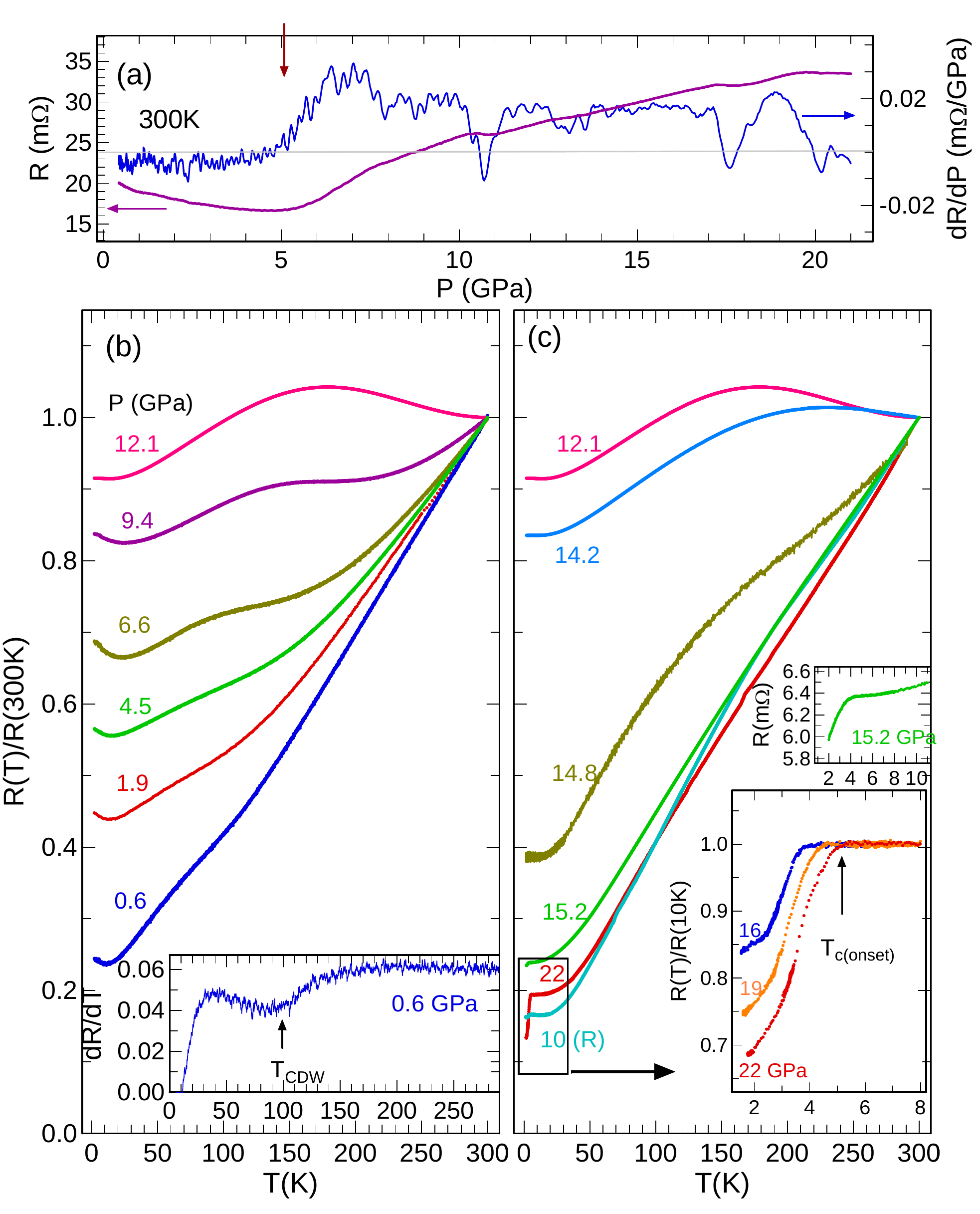}}
\caption{\label{Fig1} (Color online) (a) In-plane resistance of 1$\textit{T}$-VSe$_2$ at room temperature as a function of pressure. dR/dT plot is also shown. (b,c) R(T)/R(300K) are plotted as a function of temperature at various quasi-hydrostatic pressures. Inset in (b) shows the dR/dT plot for 0.6 GPa, the minimum at 110K indicates CDW transition temperature. Insets in (c) show R(T) at various pressures with onset of SC transition below 5K.}
\end{figure}

1$\textit{T}$-V$_{1+\delta}$Se$_2$ (with $\delta\leq$0.03) single crystals were grown by a conventional vapor transport method (with iodine as the transport agent) and characterized by x-ray diffraction, resistivity and magnetic susceptibility measurements ~\cite{misc1}. The resistance measurements were performed on the sample (dimension $\sim$120 $\mu$m x 100$\mu$m x 5$\mu$m, cut from a bigger crystal) using a standard four-probe technique (in van der Pauw configuration), with ac lock-in detection in two different high pressure arrangements. A Stuttgart version diamond anvil cell (DAC) was used under quasi-hydrostatic pressure (up to 22 GPa)~\cite{Karmakar2013}. Finely ground NaCl powder was used as the pressure medium. A pre-calibrated motorized gear was used for pressure generation in a continuous mode at $\sim$0.2 GPa/min rate to study pressure variation of resistivity at room temperature. For measurements down to 1.4K, the DAC was placed inside a KONTI-IT (Cryovac) cryostat, equipped with a homemade electromagnet coil (up to 0.5 Tesla). For high field measurements, a nonmagnetic Cu-Be DAC (M/s Easy Lab) was prepared for quasi-hydrostatic pressures (upto 15 GPa) and was inserted into a S700X SQUID magnetometer (M/s Cryogenic Ltd) to study MR and Hall resistance up to 7T field and also dc susceptibility. Pressures were measured by conventional ruby luminescence.

In Fig. 1a, room temperature in-plane resistance and its pressure derivative are shown as a function of pressure. The sample resistance initially decreases with increasing pressure upto $\sim$5 GPa (P1) where it starts increasing rapidly and at 22 GPa it becomes almost double, with two successive anomalies seen near $\sim$12 GPa and $\sim$18 GPa. Although the system remains metallic over the entire pressure range, the rapid upturn at 5 GPa showing emergence of additional scattering channels indicates significant modification of band structure near Fermi level.

\begin{figure}[tb]
\centerline{\includegraphics[width=85mm,clip]{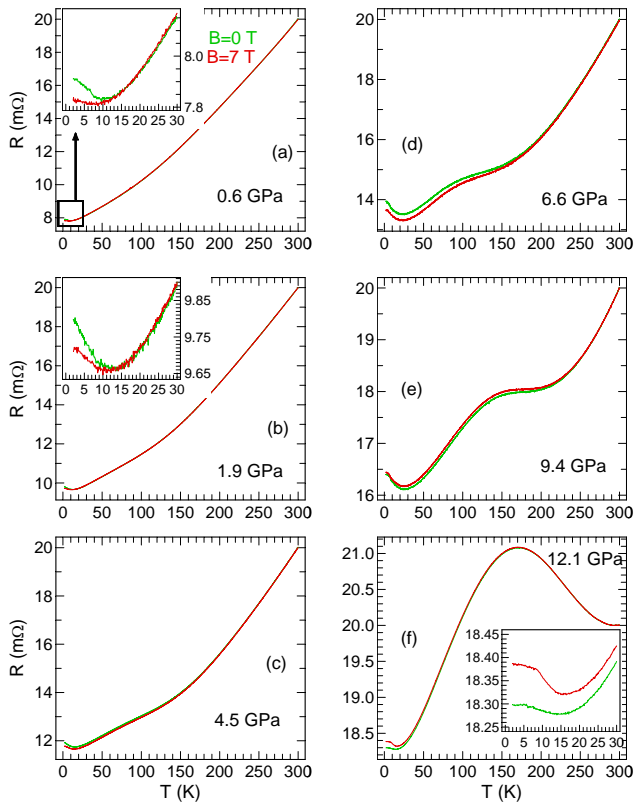}}
\caption{\label{Fig2} (Color online) (a-f) T-dependent in-plane longitudinal resistance R$_{xx}$, measured at zero field and at 7 T, at various quasi-hydrostatic $P$. Insets are magnified $R_{xx}(T)$ plots at low T below 30K.}
\end{figure}

\begin{figure}[tb]
\centerline{\includegraphics[width=90mm,clip]{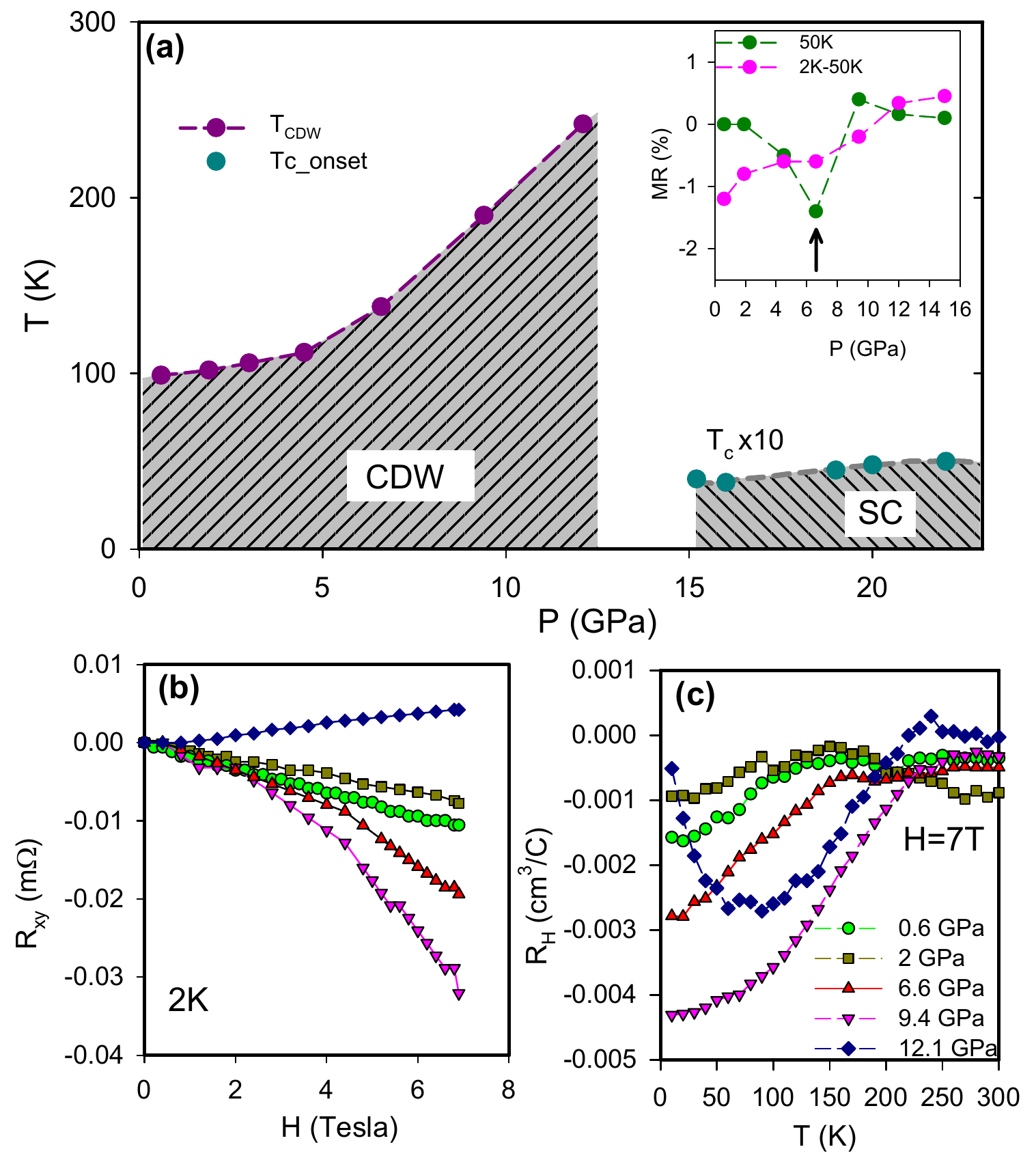}}
\caption{\label{Fig1} (Color online) (a) $P-T$ electronic phase diagram of 1$\textit{T}$-VSe$_2$ showing pressure evolution of CDW ordering T as well as SC T$_c$. Inset shows the magnetoresistance $MR(\%)$ [=($R_{7T}$-$R_{0T}$)$\times$100/$R_{0T}$] at 50K (green) and at 2K (violet, after subtracting the CDW state contribution at 50K). (b) In-plane transverse MR with zero correction (Hall resistance $R_{xy}$) at 2K as a function of applied magnetic field along c-axis and (c) Hall coefficient, $R_H$ (=$R_{xy}/H$ at 7T field) as a function of $T$ at various pressures.}
\end{figure}

\begin{figure}[tb]
\centerline{\includegraphics[width=90mm,clip]{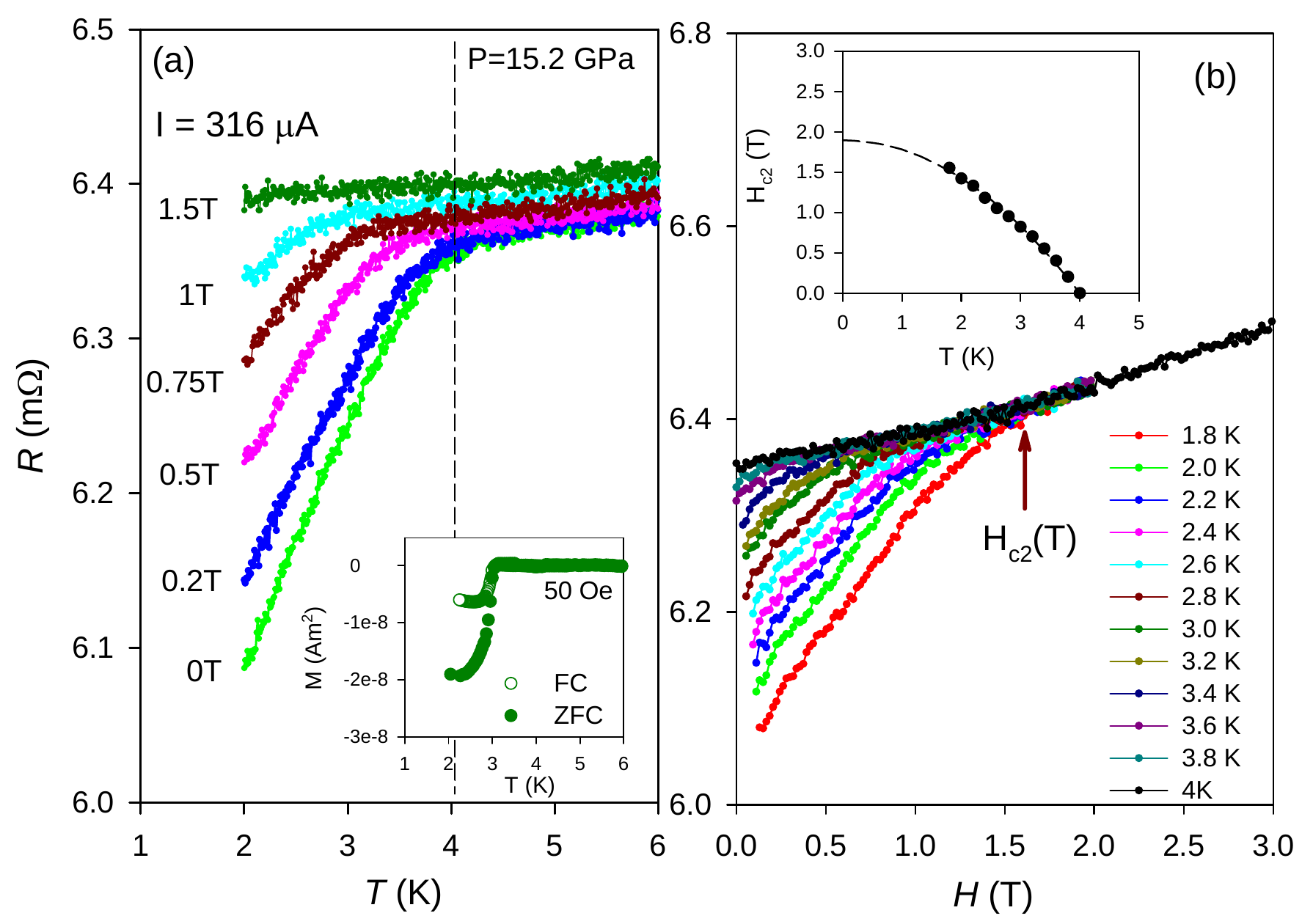}}
\caption{\label{Fig1} (Color online) (a) $R-T$ data near SC onset T$_c$ at 15 GPa under different fields up to 1.5T. Inset shows field cooled and zero-field cooled dc-susceptibility data at 50 Oe field from the sample pressurized at 15 GPa. (b) Field scanning of low temperature resistance $R(H)$ below SC T$_c$ to determine the upper critical field of the SC emerged at 15 GPa. Inset shows the H$_{c2}$(T) plots at various $T$ below T$_c$ and the GL-fitted curve.}
\end{figure}

Fig. 1(b,c) show temperature variation of the R(T)/R(300K) curves under various quasi-hydrostatic pressures up to 22 GPa. The lowest pressure (0.6 GPa) $R-T$ plot agrees well with the one at ambient pressure~\cite{misc1}. Although the CDW anomaly near 110K is not as pronounced as observed from larger single crystal, the ordering temperature can be unambiguously determined from the minimum of dR/dT curve [inset of Fig 1b]~\cite{Salvo1976,Yang2014}. A clear resistance upturn below 10K is identified as due to Kondo effect arising from the scattering of conduction electrons by interlayer V ions localized magnetic moments. The CW fitted susceptibility ~\cite{misc1} verifies a similar interlayer V ion concentration in our sample as reported in ref. ~\cite{Barua2017}, agreeing with CDW transition temperature as well as the Kondo temperature. With increasing pressure, the CDW feature in resistance is enhanced with higher ordering temperature. Also the residual resistance ratio (RRR = R$_{300K}$/R$_{10K}$) rapidly decreases from $\sim$4.4 to $\sim$1 at 12 GPa, indicating rapidly decreased metallic character. The Kondo resistance upturn is found to shift to higher T (with increased Kondo temperature) with increasing pressure up to 9.4 GPa, beyond which it decreases and eventually vanishes in the resistance data at 14.8 GPa, as can be seen in Fig 4s and Fig 5s in Supplement ~\cite{misc1}. Increased Kondo scattering at higher P is attributed to the increased effective exchange interaction between localized moments and/or enhanced density of state (DOS) at Fermi level of the host VSe$_2$~\cite{Schilling1973,Olijnyk1981}. However, as CDW order is enhanced, the metallic character decreases upon increasing P, with reduced DOS and therefore increased exchange interaction responsible for increased Kondo temperature. With further increase in pressure, CDW feature gets suppressed drastically with the RRR value at 15 GPa close to the lowest pressure [Fig. 1c]. At this pressure a significant resistance drop below 4K indicates the onset of SC transition [upper inset of Fig. 1c]. The resistance drop increases with increasing pressure with marginal increase of onset T$_c$, reaching $\sim$5K at 22 GPa. Throughout the pressure range SC transition is not complete at the lowest T (1.4K) of our set up and hence zero resistance is not achieved in four-probe resistance measurements (even from repeated loading) due to the intrinsic broad transition width [lower inset of Fig.1c]. An incomplete SC transition in this system can be attributed to the presence of interlayer Kondo impurities~\cite{Maple1976}. Also we believe a post-growth treatment of the sample may help remove strain/defects and bring the sample close to the ideal stoichiometry that may increase the superconducting volume fraction and sharpen the transition. Upon releasing pressure, the resistance drop due to onset SC is seen down to 10 GPa with enhanced T$_c$. See Fig. 7s top right panel in supplement~\cite{misc1}.

To understand the evolution of electronic structure exhibiting SC, we carried out MR and Hall measurements at high P (up to 7T field along the c-axis). Fig. 2 displays the effect of high field on the longitudinal resistance (R$_{xx}$) at various quasi-hydrostatic P. While the zero field Kondo scattering resistance upturn increases with pressure, the negative MR - the signature of scattering through magnetic moments, systematically decreases. This is due to the fact that an applied 7T field ($H< H_K=k_BT_K/\mu$ where localized moment $\mu$=2.5$\mu_B$) is able to suppress this scattering only partially causing a reduced negative MR (as discussed in Supplement~\cite{misc1}). At low pressures a small positive MR is observed above T$_{Kondo}$ as reported earlier~\cite{Barua2017}. However, at 4.5 GPa a small negative MR is seen above T$_{Kondo}$ (below T$_{CDW}$) which is much enhanced at 6.6 GPa (Fig 2d). Such negative MR can be attributed to the spin dependent scattering from the frustrated magnetic moments in triangular V lattice~\cite{Guo2014, Wong2019}. At further higher pressures (9.4 GPa) the MR reverses its sign and a positive MR reappears (See Fig 2e,f). The CDW state above T$_{Kondo}$ thus passes through a large negative MR at $\sim$6 GPa [inset of Fig 3a], indicating dramatic change in its band structure, supporting the anomalous pressure dependence of room temperature resistance at this pressure [Fig. 1]. Fig 3a displays the electronic phase diagram of 1$\textit{T}$-VSe$_2$. The CDW ordering temperature gets enhanced immediately after this band structure modification (probably by achieving better FS nesting condition), also supported by increased pressure derivative of T$_{CDW}$ at this pressure. At low pressures ($<$5 GPa), relatively smaller pressure coefficient of T$_{CDW}$ in comparison with the reported result ~\cite{Friend1978} might be due to the quasi-hydrostatic pressure medium in our measurement.

The rapid electronic structural modification is further evidenced from our high pressure Hall measurements. Fig. 3b shows pressure dependence of the in-plane transverse magnetoresistance (R$_{xy}$) measured at 2K as a function of magnetic field applied along c-axis. The negative R$_{xy}$ and its linear field dependence at low pressures agree very well with the reported results~\cite{Yang2014,Barua2017}. With increasing P initially R$_{xy}$ decreases in magnitude, indicating the enhanced carrier concentration. However, at 6.6 GPa we see a reverse trend, R$_{xy}$ starts increasing and strong non-linear field dependence is observed at high field, which is further enhanced at 9.4 GPa. The pressure-induced change in carrier behavior is also apparent from the Hall coefficient (at 7T field) plots as a function of temperature [Fig. 3c]. At 0.6 GPa, with lowering T below CDW ordering a rapid increase of negative R$_H$ is noticed due to reduced carrier concentration as a result of partial gapping near Fermi level. In the low P range carrier concentration systematically increases due to increased 3d$_{z^2}$ band-width. But for P$>$6 GPa, a rapid increase of R$_H$ is observed with large positive shift of T$_{CDW}$. The pressure-induced anomalous change in Hall coefficient is presumably due to emergence of small hole pocket at the $\Gamma$ point of the BZ. A non-linear field dependence of R$_{xy}$ also supports the two-band electronic structure.

At 12 GPa we further observe a drastic change in Hall resistance behavior; R$_{xy}$ at 2K becomes positive, indicating dominating hole contribution near Fermi level. From Fig. 2f we see that at this P, while MR near CDW still remains at a small positive value, at low T significantly large positive MR appears in the much suppressed Kondo scattering regime. At further higher pressures this positive MR persists at low temperature [Fig. 5s of ~\cite{misc1}]. A small positive MR is a common feature of non-CDW phase in pristine and pressurized layered TMDCs, originating from electron cyclotron orbital effect due to the Lorentz force~\cite{Salvo1976,Guo2014,Dutta2018}. However, the large positive MR below 10K can be due to the enhanced exchange interaction of reminiscent localized moments at this pressure. A drastic change of the RRR value above 12 GPa further supports the rapid modification of band structure near Fermi level (Fig. 1c). Moreover, T-dependent Hall coefficient at this pressure shows dramatic change in carrier concentration (Fig. 3c). Upon lowering temperature R$_H$ increases as a result of CDW ordering, but sharply decreases below 100K and changes sign below 10K. This is apparent from the positive field dependent R$_{xy}$ at 2K (Fig 3b), suggesting hole-dominated transport at this pressure. As the SC state emerges at a much higher pressure after CDW state completely gets suppressed, we can conclude that SC and CDW do not coexist. Since strong negative MR at $\sim$6GPa has been discussed in terms of spin frustration in the system, emergence of SC in the positive MR regime is speculated to originate in the vicinity of suppressed antiferromagnetic spin frustration.

In Fig. 4a, we plot the magnetic field variation of R(T) at 15.2 GPa around T$_c$. The resistance drop (with zero field onset T$_c$ $\sim$4K) is gradually lifted with increasing field, resulting in a systematic decrease in T$_c$. At a magnetic field of 1.5T the SC transition almost smears out. In all our four-probe resistance measurements we observed partial resistance drop down to 1.4K (even with lowest applied current, Fig. 6s in supplementary ~\cite{misc1}). The significant broad transition is believed to be intrinsic due to the presence of localized magnetic moments. In order to verify the SC nature, we performed DAC based dc-susceptibility measurements where significant diamagnetic drop is noticed below 3.2K [inset of Fig. 4a]. Although the bulk nature is verified, the SC shielding fraction is small (of filamentary nature) in absence of zero resistance. A field direction dependent measurement would be necessary to further understand the SC properties in this layered compound. In order to find the upper critical field, we have performed field scanning measurements at various low temperatures below T$_c$ (shown in Fig 4b). In the $T-H_{c2}$ plot [as shown in the inset of Fig. 4b], when fitted with Ginzburg-Landau (GL) formula (for BCS SC) gives H$_{c2}$(0)=1.9T, agreeing very well with typical layered TMDC SCs.

In conclusion, 1$\textit{T}$-VSe$_2$ single crystals show pressure-induced significant changes in electronic structure resulting in change in CDW ordering at 6 GPa and subsequently dramatic suppression of CDW phase above 12 GPa. Pressure-induced onset of SC transition ($\sim$4K) is observed above 15 GPa. The SC T$_c$ marginally increases with pressure up to 22 GPa. The non-dome shape feature as well as non-overlapping region with CDW phase rule out the real space co-existence of the two quantum electronic order. The measured upper critical field is of the same order as in iso-structural layered diselenide superconductors. The high pressure magnetoresistance and Hall measurements suggest successive electronic structural changes with Fermi surface modifications at 6 GPa and 12 GPa. In absence of a competitive electronic order, recently discovered intra-layer spin-fluctuation in this compound may explain the superconducting mechanism in the high pressure phase.

\begin{acknowledgments}
AKS thanks Department of Science and Technology for financial assistance. LH acknowledges DST, India (Grant No. SR/WOS-A/PM-33/2018 (G)) for funding support and Dr. Surjeet Singh for allowing the use of his crystal growth facilities.
\end{acknowledgments}

\bibliography{vse2}
\end{document}